\documentclass[prl,aps,nofootinbib,superscriptaddress,twocolumn]{revtex4}
\usepackage{graphicx} % Include figure files
\usepackage{dcolumn}  % Align table columns on decimal point
\usepackage{bm}       % bold math
\usepackage{amsmath}
\usepackage{epsfig}

\begin{document}
%\title{Spectrally-controlled excitation of dipolar sources for polarization-selective near field scattering}
\title{Experimental demonstration of linear and spinning Janus dipoles for polarisation and wavelength selective near-field coupling}
%\title{Linear and spinning dipolar sources with polarization-selective near field coupling}
%\title{Spectral tuning of dipolar sources with polarization-selective near field coupling}
%\title{Exciting dipolar sources for polarization-selective near field coupling}

% \author{M. F. Picardi$^{1,\dagger}$, M. Neugebauer$^{2,3,\dagger}$, J. S. Eismann$^{2,3,\dagger}$, P. Banzer$^{2,3*}$, F. J. Rodr\'{i}guez-Fortu\~{n}o$^{1*}$, A. V. Zayats$^1$}
% \affiliation{$^\dagger$These authors contributed equally to the work.}
% \affiliation{$^1$Department of Physics, King's College London, Strand, London, WC2R 2LS, United Kingdom}
% \affiliation{$^2$Max Planck Institute for the Science of Light, Staudtstr. 2, D-91058 Erlangen, Germany}
% \affiliation{$^3$Institute of Optics, Information and Photonics, University Erlangen-Nuremberg, Staudtstr. 7/B2, D-91058 Erlangen, Germany}

% \email[Corresponding authors: ]{Peter.Banzer@mpl.mpg.de, francisco.rodriguez_fortuno@kcl.ac.uk}
\author{M. F. Picardi}
\thanks{These authors contributed equally to the work.}
\affiliation{Department of Physics and London Centre for Nanotechnology, King's College London, Strand, London, WC2R 2LS, United Kingdom}
\author{M. Neugebauer}
\thanks{These authors contributed equally to the work.}
\affiliation{Max Planck Institute for the Science of Light, Staudtstr. 2, D-91058 Erlangen, Germany}
\affiliation{Institute of Optics, Information and Photonics, University Erlangen-Nuremberg, Staudtstr. 7/B2, D-91058 Erlangen, Germany}
\author{J. S. Eismann}
\thanks{These authors contributed equally to the work.}
\affiliation{Max Planck Institute for the Science of Light, Staudtstr. 2, D-91058 Erlangen, Germany}
\affiliation{Institute of Optics, Information and Photonics, University Erlangen-Nuremberg, Staudtstr. 7/B2, D-91058 Erlangen, Germany}
\author{G. Leuchs}
\affiliation{Max Planck Institute for the Science of Light, Staudtstr. 2, D-91058 Erlangen, Germany}
\affiliation{Institute of Optics, Information and Photonics, University Erlangen-Nuremberg, Staudtstr. 7/B2, D-91058 Erlangen, Germany}
\author{P. Banzer}
\email[]{peter.banzer@mpl.mpg.de}
\affiliation{Max Planck Institute for the Science of Light, Staudtstr. 2, D-91058 Erlangen, Germany}
\affiliation{Institute of Optics, Information and Photonics, University Erlangen-Nuremberg, Staudtstr. 7/B2, D-91058 Erlangen, Germany}
\author{F. J. Rodr\'{i}guez-Fortu\~{n}o}
\email[]{francisco.rodriguez_fortuno@kcl.ac.uk}
\affiliation{Department of Physics and London Centre for Nanotechnology, King's College London, Strand, London, WC2R 2LS, United Kingdom}
\author{A. V. Zayats}
\affiliation{Department of Physics and London Centre for Nanotechnology, King's College London, Strand, London, WC2R 2LS, United Kingdom}

%\date{\today}

\begin{abstract} %<150 words
The electromagnetic field scattered by nano-objects contains a broad range of wave vectors and can be efficiently coupled to waveguided modes. The dominant contribution to scattering from subwavelength dielectric and plasmonic nanoparticles is determined by electric and magnetic dipolar responses. Here, we experimentally demonstrate spectral and phase selective excitation of Janus dipoles, sources with electric and magnetic dipoles oscillating out of phase, in order to control near-field interference and directional coupling to waveguides. We show that by controlling the polarisation state of the dipolar excitations and the excitation wavelength to adjust their relative contributions, directionality and coupling strength can be fully tuned. Furthermore, we introduce a novel \emph{spinning} Janus dipole featuring cylindrical symmetry in the near and far field, which results in either omnidirectional coupling or noncoupling. Controlling the propagation of guided light waves via fast and robust near-field interference between polarisation components of a source is required in many applications in nanophotonics and quantum optics. 

%To date, such versatile control has enabled the routing of light emitted from a dipolar source into any desired direction in waveguides. A dipolar electromagnetic source that presents a face-selective coupling to guided modes was theoretically predicted recently. This so-called Janus dipole will or will not excite modes, depending on which of its two ``faces'' - coupling or noncoupling one - is shown to the waveguide. %This selective coupling means that, when the Janus dipole is placed between two identical waveguides, it will excite a guided mode inside one of the two and, at the same time, leave the other one unexcited. In this work we experimentally realize a Janus dipole and measure its near-field scattering behaviour. and observe, for a given polarization, an omnidirectional noncoupling behaviour, corresponding to the impossibility of the source to excite a mode propagating in \emph{any} direction. 
% [[Flipping the polarization will lead, in turn, to a full omnidirectional coupling behaviour, revealing the two faces of the Janus dipole.]] % But we do not measure this. This can be very confusing to a reader, because this dipole is coupling for p and non-coupling for s.

%{\bf One sentence summary:} %< 150 characters 
%We experimentally demonstrate omnidirectional coupling and non-coupling states via spectral selective excitation of electric and magnetic dipoles in nano-objects.
%We experimentally demonstrate polarization selective near-field coupling via spectrally controlled interference of linear or circularly polarised electric and magnetic dipoles.

\end{abstract}

\maketitle

Scattered fields from plasmonic and dielectric nanostructures contain a broad range of wavevectors  \cite{bohren1983absorption} which make them suitable for efficient coupling to waveguided modes, underpinning applications in photonic data manipulation, quantum technologies and precision metrology. The multipole expansion of these fields can, in many cases, be limited to the lowest-order multipoles. In particular, the optical response of small plasmonic nanoparticles can be approximated by the lowest electric dipole scattering contribution, while for high-index dielectric nano-objects the dominating contributions are electric and magnetic dipoles \cite{evlyukhin2012demonstration,wozniak2015selective,eismann2018exciting,zywietz2014laser,wei2014control}. Controlling the relative amplitudes and phases between the induced dipoles allows for complete engineering of the polarisation of the nanostructured sources \cite{wozniak2015selective,eismann2018exciting,neugebauer2016polarization}, leading to interesting near- and far-field behaviours \cite{wei2014control}. For example, spin-momentum locking in guided modes \cite{bliokh2015quantum} makes it possible to achieve directionality of light using circularly polarised dipolar sources \cite{rodriguez2013near,o2014spin,neugebauer2014polarization,kapitanova2014photonic,petersen2014chiral} with a broadband robust behaviour, which, in principle, is switchable at ultrafast speeds limited only by the light field oscillation cycle. This has led to numerous applications in polarimetry \cite{fortuno2014universal,espinosa2017chip, neugebauer2015measuring}, quantum optics \cite{coles2016chirality,le2015nanophotonic,lodahl2017chiral}, and optical manipulation \cite{hayat2015lateral,rodriguez2015lateral,sukhov2015dynamic}, amongst many others. 

While circularly polarised electric dipoles, comprised of two orthogonal linear electric dipoles oscillating with a phase difference of $\pm \pi/2$, excite unidirectionally $p$-polarised waveguide modes, circular magnetic dipoles can be used to excite $s$-polarised modes \cite{picardi2017unidirectional}. By superimposing electric and magnetic dipole contributions, additional directionalities can be achieved via their interference \cite{neugebauer2016polarization}. In this context, the Huygens dipole is a combination of orthogonally oriented, in-phase, electric $\mathbf{p}$ and magnetic $\mathbf{m}$ dipoles, which fulfil Kerker's scattering condition, $p= m/c$, with $c$ being the speed of light. This source is well known to exhibit directionality in the far-field \cite{neugebauer2016polarization,evlyukhin2015resonant,staude2013tailoring} and is employed in reflectionless dielectric metasurfaces \cite{kuznetsov2016optically,liu2017huygens}. Moreover, when the electric and magnetic dipoles are perpendicular to each other, like in the Huygens dipole, but $\pm \pi/2$ out of phase, the resulting source is the so-called Janus dipole, which has been recently predicted \cite{picardi2018janus}. It earns its name from the different behaviour observed depending on which side of this source faces a nearby waveguide. One face will couple to guided modes, while the opposite one will show a complete absence of coupling. This behaviour is reversed by flipping the polarisation of the dipole, i.e. switching between the two faces. Janus, Huygens, and circularly polarised dipoles were identified as the three elemental dipolar sources for directional mode excitation in planar geometries \cite{picardi2018janus,picardi2018not}. All these sources are based on the same fundamental principles of near field interference, provide broadband operation, and can be controlled at ultrafast speeds by switching the polarisation of the excitation light.  

\begin{figure*}[htbp]
\includegraphics[width=\linewidth]{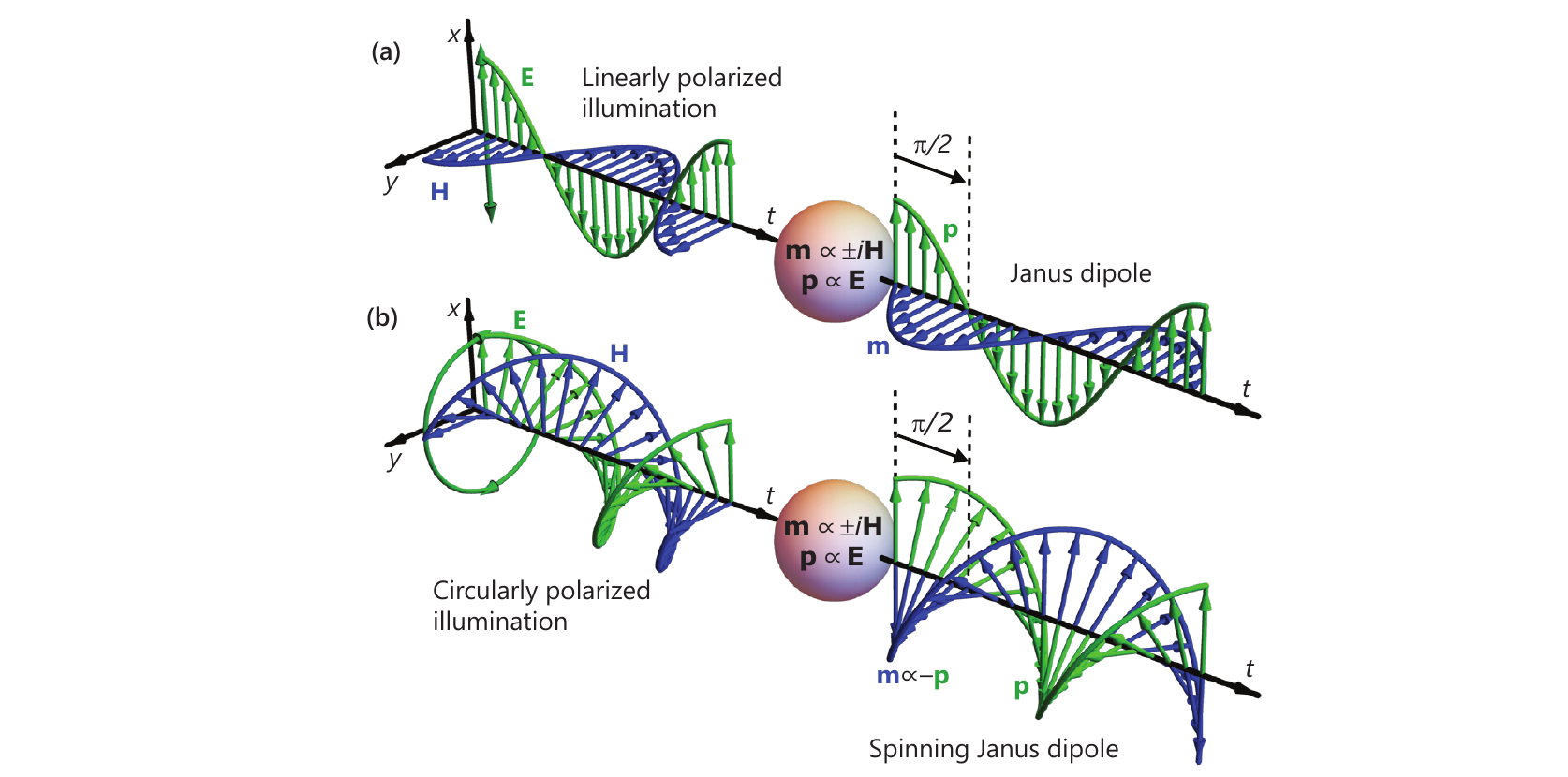}
\caption{\textbf{Linear and spinning Janus dipoles.} A nanoparticle whose electric and magnetic polarizabilities have a fixed phase difference at a given wavelength scatters light like a dipolar source with electric and magnetic dipole moments featuring a phase difference determined by the intrinsic polarizabilities. Left: incident $\mathbf{E}$ and $\mathbf{H}$ fields as a function of time. Right: dipole moments $\mathbf{p}$ and $\mathbf{m}$ of the nanoparticle as a function of time. When the phase difference between the two polarizabilities is $\pi/2$, (a) the nanoparticle under linearly polarised plane wave illumination will scatter like a linear Janus dipole. (b) The same nanoparticle under circularly polarised plane wave illumination will scatter like a spinning Janus dipole (i.e., electric and magnetic fields rotate in the same sense and are oriented antiparallel at all times). An additional global phase-delay between the excitation fields and the resulting dipoles is omitted in the sketch.}\label{fig:fig1_new}
\end{figure*}

Here we experimentally achieve wavelength-selective excitation of Janus dipole sources, with their directional coupling properties, by tailoring the near-field interference between electric and magnetic dipole moments induced in dielectric nanoparticles. We show that by tuning the polarisation state of the excited dipoles and the excitation wavelength to adjust their relative contributions, different dipolar sources can be realized, such as the linear Janus dipole. In addition, we discuss and experimentally prove the possibility of achieving omnidirectional coupling or noncoupling with a novel spinning Janus dipole. 
\section{Results}
\subsection{Janus dipole excited in a nanoparticle}

A dipolar source can be realized experimentally by illuminating any small nanostructure scattering in the lowest order Mie regime \cite{neugebauer2014polarization,neugebauer2016polarization,evlyukhin2015resonant}. Simultaneous electric and magnetic dipolar excitations will be achieved if it has both electric and magnetic polarizabilities different from zero \cite{evlyukhin2012demonstration,staude2013tailoring,kuznetsov2016optically,zywietz2014laser}. Plane-wave illumination conveniently provides orthogonal electric $\mathbf{E}$ and magnetic $\mathbf{H}$ fields, matching the $\mathbf{p}$ and $\mathbf{m}$ dipole moment directions required for the linear Janus dipole. However, the orthogonal fields of plane waves are always in phase. In order to obtain a Janus source whose electric and magnetic dipole moments are phase shifted, we can exploit the intrinsic wavelength-dependent phase difference between the electric and magnetic polarizabilities of the particle \cite{neugebauer2014polarization,neugebauer2016polarization}. When this phase difference equals $\pm\pi/2$, and the amplitudes of the electric and magnetic dipole moments are comparable, a Janus dipole is achieved (Fig.~\ref{fig:fig1_new}(a)).
\begin{figure}[htbp]
\includegraphics[width=\linewidth]{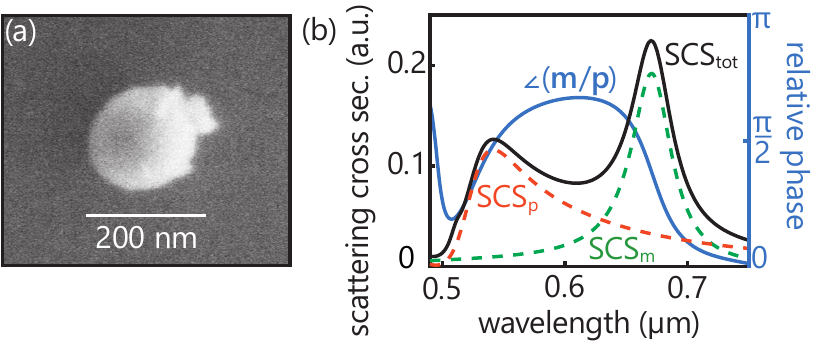}
\caption{\textbf{Nanoparticle properties.} (a) SEM image of a spherical Si particle with core radius $r =84~\mathrm{nm}$ and oxide shell thickness $s=4~\mathrm{nm}$. (b) Total scattering cross-section (solid black line) and relative phase between the resonances (solid blue line) calculated using Mie theory for the nanoparticle in (a). The dashed orange and green lines indicate the electric and magnetic dipole contributions, respectively.}\label{fig:fig1}
\end{figure}
High-index dielectric nanoparticles, such as silicon particles, are suitable for this purpose since they possess both electric and magnetic Mie resonances \cite{wozniak2015selective,neugebauer2016polarization}. Moreover, depending on the size of the particle, higher order multipole resonances can be safely neglected in the visible spectrum \cite{evlyukhin2012demonstration}.
%In Fig.~\ref{fig:fig1}(a), the scanning electron micrograph of the silicon nanoparticle investigated experimentally is shown; in (b) the corresponding scattering cross-sections retrieved from Mie scattering theory are plotted. 
%It is important that the nanoparticle is isolated to avoid undesired interference due to neighbouring sources.
By tuning wavelength and polarisation of the illumination, we can select amplitudes, direction, and phase difference of the electric and magnetic dipole moments in the nanoparticle, making it the ideal candidate to experimentally realise a Janus dipole source.
% explanation of s-pol vs p-pol behaviour of Janus dipole added here

The unique coupling behaviour of a Janus dipole with a waveguide is closely related to the reactive power of the evanescent tails in the mode being excited \cite{picardi2018janus}. Reactive power is the vector $\mathrm{Im}\left\lbrace\mathbf{E}^*\times\mathbf{H}\right\rbrace$, i.e., the imaginary part of the Poynting vector. The coupling or noncoupling behaviour of the Janus dipole depends on whether the corresponding vector quantity $\mathrm{Im}\left\lbrace\mathbf{p}^*\times\mathbf{m}\right\rbrace$ of the source is pointing in the same or in opposite direction to the reactive power of the mode. This gives rise to its two faces. The direction of the reactive power of an evanescent wave depends on its polarisation \cite{picardi2018janus,wei2018directional}: $s$-polarised waves (also called transverse electric, with no electric field component in the direction of propagation) have a reactive power which points in the direction of the evanescent decay, while the reactive power of $p$-polarised modes (transverse magnetic) is opposite to the direction of decay. Therefore, the definition of coupling and non-coupling faces of a Janus dipole depends on the polarisation of the excited mode. In this work, we experimentally generate both a linear and a spinning Janus dipole with $\mathrm{Im}\left\lbrace\mathbf{p}^*\times\mathbf{m}\right\rbrace$ pointing towards a nearby medium of higher optical density (glass with a refractive index of 1.5), resulting in preferred $p$-polarized and strongly suppressed $s$-polarized evanescent coupling between the dipole and the medium. We observe this behaviour by measuring the angular spectrum of the sources in the glass half-space (similar to the measurements shown in \cite{neugebauer2014polarization,neugebauer2016polarization}, see SM for details).
%, showing a complete circle of zeros in $k$-space for its $s$-polarized component, which corresponds to a 360-degree absence of coupling, i.e. the light emitted by our source cannot excite an $s$-polarized guided mode in any direction. 
%The results are compared with simulations, which agree very well.
% Finally, we consider a \emph{rotating} Janus dipole which shows a peculiar behaviour: its $p$- and $s$-polarized components are cylindrically symmetric: the first will excite modes in \emph{any} direction with the same amplitude, the second will be noncoupling, in the same way. % I would not add this until the end, as it is not really a summary

\subsection{Experimental verification}

%Before discussing the measurement of a novel spinning Janus dipole, 
Because of the small distance between the dipolar source and the substrate, both the propagating and evanescent wave-vector components of the source can couple into propagating waves inside the glass, which can be measured. We are interested in the emission that corresponds to evanescent fields in free-space, responsible for the near field directionality of the Janus dipole, with $k_{t}/k_0 > 1$, where $k_{t}$ is the transverse wavevector perpendicular to the optical axis ($z$) and $k_0$ is the wavenumber in free space. Although the amplitude of the measured spectrum will be a modified version of the near field spectrum of the isolated Janus source in free space, the difference can be understood via a multiplicative transfer function that accounts for the polar angle dependence of the Fresnel transmission coefficients through the high index substrate interface. Therefore, any zeroes in the angular spectrum of the free-space source will also be present in the measured angular spectra. This follows directly from the conservation of transverse momentum. The arrangement of zeroes in the spectra are a clear signature of a Janus dipole (see SM). For instance, a Janus dipole with $p_x/m_y = -iR/c$, where $R$ is a normalized measure of the ratio of electric to magnetic components, shows zero amplitude for the $s$-polarised evanescent components with $k_t = k_0 \sqrt{R^2+1}$, on its noncoupling side $(z>0)$, owing to the destructive interference between the electric and magnetic dipole fields after their superposition. A Janus dipole with $p_x/m_y = -i/c$ would then lead to a ring of zero intensity at the transverse $k$-vector $k_t=\sqrt{2}k_0$ \cite{picardi2018janus}. However, this would exceed the angular range of our experimental setup, so our ideal Janus dipole condition is $p_x/m_y =-i 0.75/c$, optimized for $k_t=1.25 k_0$.

For our experiment, we place an individual silicon nanosphere (diameter $176\,\text{nm}$) on a glass substrate [see Fig~\ref{fig:fig1}(a)] on the optical axis of a linearly $x$-polarised Gaussian beam (focused with an effective NA of 0.5) used for excitation. For this configuration, owing to the linearly polarised illumination, we  excite an $x$-polarised electric dipole $p_x$ and a $y$-polarised magnetic dipole $m_y$. We can then control the amplitudes and the relative phase between both dipole moments by selecting the wavelength of the excitation field. Between the magnetic and electric dipole resonances [Fig~\ref{fig:fig1}(b)], we expect two wavelengths for which the relative phase between $m_y$ and $p_x$ is close to $\pi/2$ (Janus dipole condition) with both dipole amplitudes being of comparable strength. Due to the presence of the substrate, these will be slightly different from the ones predicted by free-space Mie theory. Nonetheless, the free-space scattering cross section provides for a range within which the wavelength can be fine-tuned experimentally. We then measure the intensity distribution in the back focal plane (BFP) of an oil immersion objective ($\mathrm{NA} = 1.3$) placed below the glass substrate, capturing the near and far-field parts of the angular spectrum of the Janus dipole for $0.6<k_t/k_0<1.3$. The angular range below an NA of 0.6 is also collected but discarded, because it contains the transmitted input beam. The collected spectrum is analysed with a linear polariser to retrieve its $s$- and $p$-polarisation components.

\begin{figure*}[htbp]
\includegraphics[width=170mm]{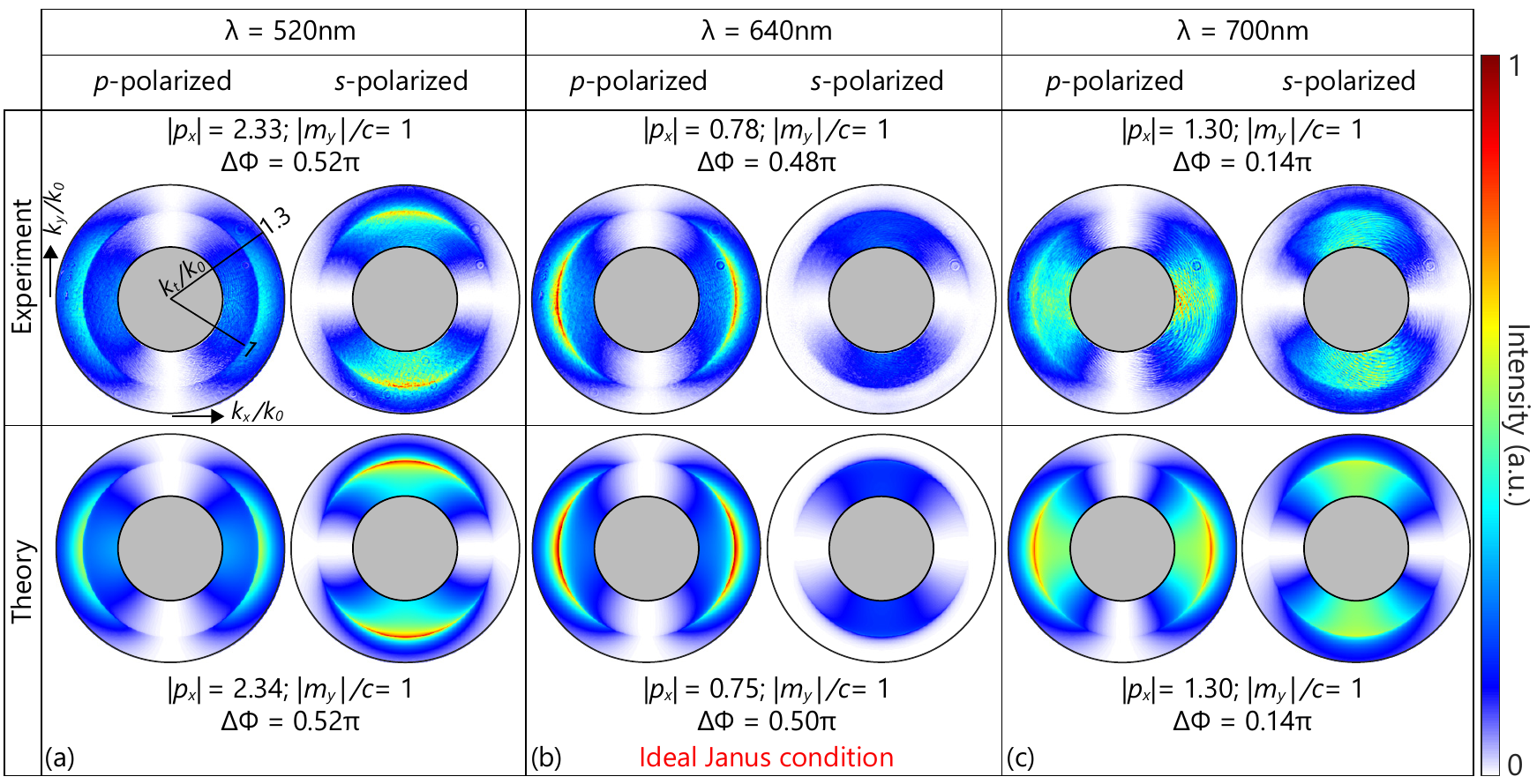}
\caption{\textbf{Near-field scattering of dipolar sources and their spectral dependence.} Measured (top) and calculated (bottom) BFP intensities of $s$- and $p$-polarised near-field angular distributions for (a) $\lambda = 520\,\text{nm}$, (b) $\lambda = 640\,\text{nm}$, (c) $\lambda = 700\,\text{nm}$ corresponding to different relative contributions of electric and magnetic dipoles and their different relative phases. For $\lambda = 640\,\text{nm}$, the Janus condition ($|p_x/m_y|=0.75/c$ and $\Delta\Phi=\mathrm{arg}(m_y/p_x)=\pi/2$) is theoretically achieved. It can be seen that the $s$-polarised component of the light scattered by the Janus dipole presents a full ring of zero intensity at the locations $ k_{x}^2+k_{y}^2=k_t^2=k_{0}^2(R^2+1) $ corresponding to the noncoupling condition for a given \emph{amplitude} of the wavevector $k_{t}$, in \emph{any} direction. In (a) and (c), the dipole moments and phase difference used in the theoretical plots are the same as the experimental ones, while in (b) the simulation shows the ideal Janus dipole, with $|p_x/m_y|=0.75/c$ and $\Delta\Phi=\pi/2$.}\label{fig:fig2}
\end{figure*}

Figure~\ref{fig:fig2} shows the results of the measurements together with the calculated dipoles, obtained for three different wavelengths of the illumination: (a) $\lambda=520\,\text{nm}$, (b) $\lambda=640\,\text{nm}$, and (c) $\lambda=700\,\text{nm}$. We notice a striking agreement between the measured data and the numerical calculations. At $\lambda = 640\,\text{nm}$ we are very close to the linear Janus dipole condition [Fig.~\ref{fig:fig2}(b)], for which the electric and magnetic dipole moments have a relative phase close to $\pi/2$ and an amplitude ratio $|p_x/m_y|\approx 0.75/c$.  A ring of zero amplitude outside the light cone (corresponding to near fields) in the $s$-polarised angular spectrum of Fig.~\ref{fig:fig2}(b) is a clear signature of the noncoupling face of the Janus dipole. In fact, our measurements reveal that the amplitude of its angular spectrum is zero for a circle with transverse wavevector $k_t=1.25~k_0$ as analytically expected. Hence, if placed in close proximity to a waveguide supporting an $s$-polarised mode with this or similar propagation constant, this source will not be able to excite it in \emph{any} direction due to a momentum mismatch. The $p$-polarised component, on the other hand, is non-zero everywhere except for the $k_x = 0$ line. The source will excite $p$-polarised modes in all directions except for the $\pm y$-direction.
These are trivial zeroes because they are the result of a polarisation mismatch: the dipole has components $p_x$ and $m_y$, but $p$-polarised modes propagating parallel to the $y$-direction do not feature the corresponding $E_x$ and $H_y$ field components to couple to.
% However, reversing the polarization of the source (flipping the face of the Janus dipole), the angular spectra of $s$- and $p$-polarized components swap and rotate by 90 degrees, which means that the $p$-polarized component will not excite modes in any direction (full circle of zeros), while the $s$-polarized component will now be able to couple to modes propagating in any direction other than $k_y=0$ (See Appendix for the full mathematical description of the angular spectrum).\\ % I would not add this because, unfortunately, we do not measure it. Also it seems to me uneccessarily complex

Figure~\ref{fig:fig2}(a) and (c) show the angular spectra obtained at two other wavelengths, $520\,\text{nm}$ and $700\,\text{nm}$ respectively, for which the Janus condition is not fulfilled and, therefore, the aforementioned feature of noncoupling can not be achieved. 
%From the electric and magnetic spectral response of the particle, Fig~\ref{fig:fig1}(c), we can predict the dipole moments of the nanoparticle at those wavelengths. 
From the angular spectra measurements, we can retrieve the corresponding dipole moments induced in the nanoparticle at those wavelengths \cite{eismann2018exciting}.
For $\lambda = 520~\mathrm{nm}$, the amplitude of the magnetic dipole moment is significantly smaller than the electric one ($|p_x/m_y|\approx 2.3/c$). Hence, even if the phase between both of them is close to $\pi/2$, the destructive interference condition happens at transverse wave-vectors $k_t/k_0 \gg \mathrm{NA}$, well above the available numerical aperture in the experiment. In the measured angular region, the electric dipole behaviour will be dominant and the nanoparticle will scatter like an electric dipole polarised along $x$. On the other hand, for $\lambda=700\,\text{nm}$, the amplitudes of the two dipole moments are comparable $|p_x/m_y| \approx 1.3/c$ but the phase between the two is almost zero $\Delta\Phi=0.14$.
\subsection{Spinning Janus dipole}
\begin{figure}[htbp]
\includegraphics[width=\linewidth]{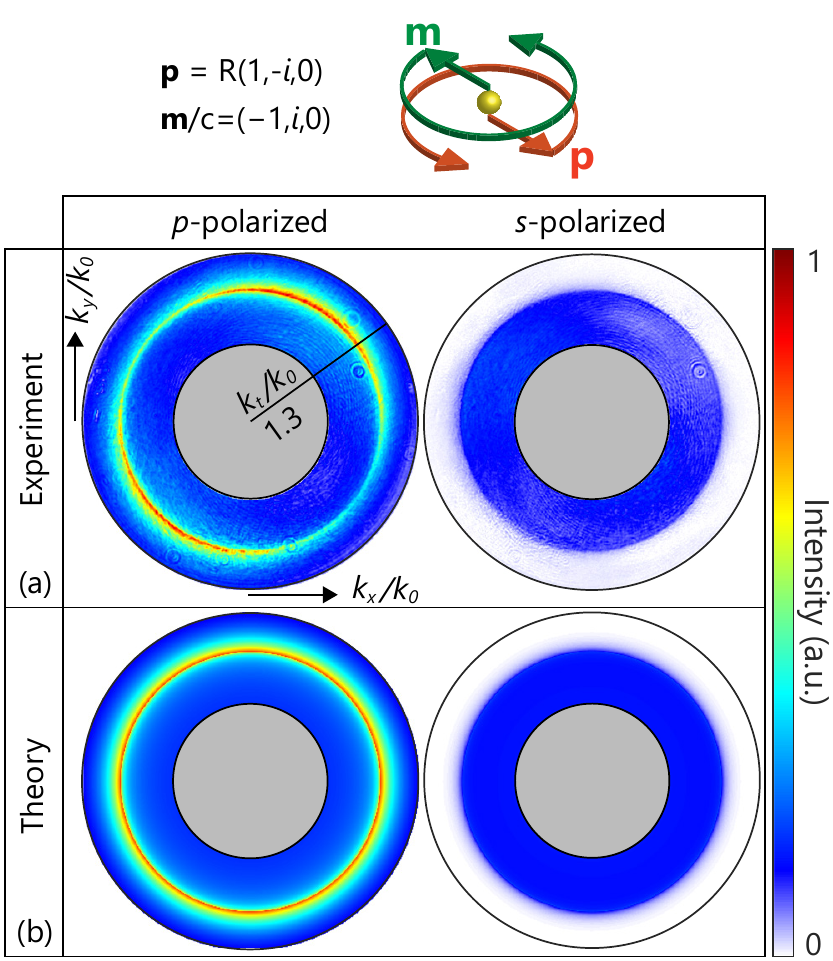}
\caption{\textbf{Spinning Janus dipole}. Measured (a) and calculated (b) BFP intensities of the $p$- and $s$- polarised scattering from a spinning Janus dipole. The retrieved experimental dipole moments are $\mathbf{p}=(0.61-0.11i,-0.01-0.75i,0)$ and $\mathbf{m}/c=(-1, 0.30+0.84i, 0)$, while the theoretical ones are $\mathbf{p}=0.75(1, -i, 0)$ and $\mathbf{m}/c=(-1, i, 0)$. The slight asymmetry in the experimental results can be attributed to the experimental dipole moments not matching exactly the ideal ones for which the spinning Janus dipole is cylindrically symmetric.}\label{fig:fig3}
\end{figure}

As a consequence of the linearly polarised illumination, $\mathbf{p}$ and $\mathbf{m}$ are always pointing along $x$ and $y$, respectively [Fig.\ \ref{fig:fig1_new}(a)]. This is responsible for the lines of zero amplitude ($k_y = 0$ for $s$-polarised light and $k_x=0$ for $p$-polarised light) clearly visible in all angular spectra in Fig.~\ref{fig:fig2}. These zeroes are caused by a polarisation mismatch between the dipole and the modes, as described above, rather than the destructive interference between $\mathbf{p}$ and $\mathbf{m}$ characteristic for the Janus dipole. 
A way to remove these trivial lines of zero amplitude in the spectra is via illumination with circularly polarised light. This should result in the excitation of $\mathbf{p}$ and $\mathbf{m}$ with the same time-dependence as the illuminating $\mathbf{E}$ and $\mathbf{H}$ fields, but with a phase delay corresponding to $\pi/2$, as a direct consequence of the particle's response [Fig.\ \ref{fig:fig1_new}(b)]. This induces electric and magnetic dipoles which are circularly polarised, spinning together in the $xy$ plane, but being oriented anti-parallel at all times, such that  $\mathbf{p} = (1, -i, 0)$ and $\mathbf{m}/c = -\mathbf{p}/R$. This constitutes a novel ``spinning'' Janus dipole, with a non-zero associated vector $\mathrm{Im}\lbrace\mathbf{p}^*\times\mathbf{m}\rbrace$ directed along $+z$, towards the substrate, as required for non-coupling to $s$-polarised modes. The angular spectrum intensity of this dipole is evidently rotationally symmetric, showing no polarisation mismatch to modes in any direction. The stark contrast between $p$-polarised coupling and $s$-polarised non-coupling in the evanescent region is even clearer in the experiment. The full ring of zeroes is caused purely by the interference of $\mathbf{p}$ and $\mathbf{m}$, characteristic for the Janus dipole [Fig.~\ref{fig:fig3}]. The source couples to $p$-polarised evanescent waves in all directions, while it does not couple to $s$-polarised evanescent waves with a fixed $k_t>k_0$, at any angle. This behaviour would be reversed for an opposite sign of the $\pi/2$ phase difference between the electric and magnetic polarizability induced by the nanoparticle. In this case, the electric and magnetic dipoles induced are spinning parallel to each other and the source is noncoupling for $p$-polarised modes. 

\section{Discussion}

% The possibility to determine the coupling of a source to guided modes in any direction by means of its polarization opens unprecedented avenues for nanoscale routing and selective illumination. Measuring the Janus dipole's near field we have confirmed the theoretical prediction of a two-faced source with opposite, side-dependent behaviours. The simplicity of the geometry of the source, an illuminated silicon nanosphere, makes it also suitable for integration in silicon-based photonics circuits. The cylindrical symmetry of the near field radiation of the rotating Janus dipole had not yet been observed in any dipole-like source. This could potentially make the rotating Janus dipole the most likely choice for a  fully omnidirectional antenna in the near field. The striking difference between the $s$- and $p$-polarized spectra could trigger substantial advances in polarization-selective excitation of guided modes and 
In conclusion, the experimental measurement of the Janus dipole confirms the theoretical predictions of a source with a polarisation-dependent omnidirectional absence of coupling to evanescent waves. This adds to the already widely used circular and Huygens dipoles as an extra source with a polarisation-controllable near field. The striking agreement between the dipoles obtained by the scattering from the silicon nanoparticle and the theoretical point sources validates the effectiveness of the utilized dipolar approximation. Moreover, the sensitivity of the response to the illumination parameters leaves room for applications in which a different phase and amplitude ratio between the dipole components may be required. This includes, but is not limited to, guided modes with different transverse wavevectors, which can be matched to the source by properly tuning the dipole components. This experimental demonstration highlights the feasibility of the Janus source, paving the way towards novel applications in nanophotonics, quantum information and plasmonics, which might include the Janus dipole and its spinning version.

\section{Acknowledgements}
This work was supported by European Research Council Starting Grant ERC-2016-STG-714151-PSINFONI, EPSRC (UK) and ERC iCOMM project (789340). A.Z. acknowledges support from the Royal Society and the Wolfson Foundation. All the data supporting this research are provided in full in the results section and Supplementary Materials.

\bibliographystyle{unsrt}
% * <a.zayats@kcl.ac.uk> 2018-08-21T10:32:23.042Z:
% 
% > %\bibliographystyle{unsrt}
%  30 references maximum in the main text.
% 
% ^ <michela.picardi@kcl.ac.uk> 2018-09-07T13:51:05.724Z:
% 
% done
%
% ^.

\bibliography{exp_janus.bib}

\end{document}